\documentclass[conference]{IEEEtran}

\usepackage{float}
\IEEEoverridecommandlockouts

\usepackage[authoryear,round]{natbib}
\usepackage[hidelinks]{hyperref}
\usepackage{xcolor}
\usepackage{pgffor} 

\makeatletter
\newcommand{\citepblueyear}[1]{%
  (%
  \foreach \x [count=\i from 1] in {#1} {%
    \ifnum\i>1; \fi%
    \citeauthor{\x}~\textcolor{blue}{\citeyear{\x}}%
  }%
  )%
}
\makeatother

\usepackage{braket}
\usepackage{amsmath,amssymb,amsfonts}
\usepackage{algpseudocode} 
\usepackage{algorithm}
\usepackage{graphicx}
\usepackage{subcaption}
\usepackage{textcomp}
\usepackage{titlesec}
\usepackage{url}
\titleformat{\section}
  {\normalfont\Large\bfseries\filright}
  {\thesection}
  {0pt}
  {}

\titleformat{\subsection}
  {\normalfont\large\bfseries}
  {\thesubsection}
  {1em}
  {}

\def\BibTeX{{\rm B\kern-.05em{\sc i\kern-.025em b}\kern-.08em
    T\kern-.1667em\lower.7ex\hbox{E}\kern-.125emX}}

\renewcommand{\thesection}{\arabic{section} }
\makeatletter
\def\thesubsection{\thesection.\arabic{subsection}}
\def\p@subsection{}
\makeatother

\begin{document}

\title{Differentiable Architecture Search for Adversarially Robust Quantum Computer Vision\thanks{Published in \textit{Quantum Machine Intelligence} (2026) 8:2. DOI: https://doi.org/10.1007/s42484-026-00353-0}}
\author{
Mohamed Afane\textsuperscript{1*} \quad Quanjiang Long\textsuperscript{1} \quad Haoting Shen\textsuperscript{2} \quad Ying Mao\textsuperscript{1} \\ Junaid Farooq\textsuperscript{3} \quad Ying Wang\textsuperscript{4} \quad Juntao Chen\textsuperscript{1*}\\
\textsuperscript{1}Fordham University \quad
\textsuperscript{2}Zhejiang University \quad
\textsuperscript{3}University of Michigan-Dearborn \quad
\textsuperscript{4}Stevens Institute of Technology \\
}

\twocolumn[
\begin{@twocolumnfalse}

\maketitle

\vspace{-1em}
{$^{\ast}$Correspondence: M. Afane: mafane@fordham.edu; J. Chen: jchen504@fordham.edu\\
\small $^{\dagger}$Published in \textit{Quantum Machine Intelligence} (2026) 8:2. DOI: https://doi.org/10.1007/s42484-026-00353-0\\}

\vspace{0.5em}
\noindent\textbf{Abstract} \\
Current quantum neural networks suffer from extreme sensitivity to both adversarial perturbations and hardware noise, creating a significant barrier to real-world deployment. Existing robustness techniques typically sacrifice clean accuracy or require prohibitive computational resources. We propose a hybrid quantum-classical Differentiable Quantum Architecture Search (DQAS) framework that addresses these limitations by jointly optimizing circuit structure and robustness through gradient-based methods. Our approach enhances traditional DQAS with a lightweight Classical Noise Layer applied before quantum processing, enabling simultaneous optimization of gate selection and noise parameters. This design preserves the quantum circuit's integrity while introducing trainable perturbations that enhance robustness without compromising standard performance. Experimental validation on MNIST, FashionMNIST, and CIFAR datasets shows consistent improvements in both clean and adversarial accuracy compared to existing quantum architecture search methods. Under various attack scenarios, including Fast Gradient Sign Method (FGSM), Projected Gradient Descent (PGD), Basic Iterative Method (BIM), and Momentum Iterative Method (MIM), and under realistic quantum noise conditions, our hybrid framework maintains superior performance. Testing on actual quantum hardware confirms the practical viability of discovered architectures. These results demonstrate that strategic classical preprocessing combined with differentiable quantum architecture optimization can significantly enhance quantum neural network robustness while maintaining computational efficiency.


\vspace{1em}  
\end{@twocolumnfalse}
]

\vspace{1em}

\section{Introduction}

Recent years have witnessed a surge of research in quantum machine learning propelled by advances in quantum computing hardware and novel algorithmic techniques \citepblueyear{feynman2018simulating, peruzzo2014variational, hadfield2019quantum}. Among the various applications, quantum computer vision (QCV) has emerged as a particularly active area, where quantum neural networks (QNNs) have been explored and shown competitive performance on image classification tasks \citepblueyear{dang2018image, tian2023recent, afane2025atp}. While these models can potentially exploit quantum superposition and entanglement for richer feature representations, their practical deployment often faces challenges related to inexpressibility, computational overhead, and robustness \citepblueyear{abbas2021power}.

One major hurdle in QNN development is discovering effective circuit architectures. Conventional approaches like evolutionary algorithms and reinforcement learning can navigate this space but are typically computationally expensive. In contrast, differentiable quantum architecture search (DQAS) provides a more efficient framework by using gradient-based methods to optimize model parameters.
\citepblueyear{martyniuk2024quantum, wu2023quantumdarts, zhang2022differentiable, chen2025introduction, chen2025differentiable}. This strategy has shown promising results in tasks like optimizing combinatorial problems \citepblueyear{hadfield2019quantum}, material science \citepblueyear{sajjan2022quantum}, quantum optimization and image-based applications \citepblueyear{chen2024differentiable, sun2023differentiable}. 

Besides accuracy, robustness remains a critical concern for real-world scenarios in QCV. Small adversarial perturbations or inherent hardware noise can drastically degrade performance, especially given that the delicate states in QNNs are highly sensitive to interference. Techniques that increase circuit entanglement \citepblueyear{maouaki2024robqunns} can make decision boundaries more complex but may demand excessive quantum resources and slow down training \citepblueyear{lu2020quantum}. Other defensive methods such as randomized encoding or quantum error correction \citepblueyear{gong2024enhancing} often come at the cost of reduced clean accuracy or impose steep computational overheads that limit scalability \citepblueyear{martyniuk2024quantum}.

\begin{figure}[t]
\centering
\includegraphics[width=1\columnwidth]{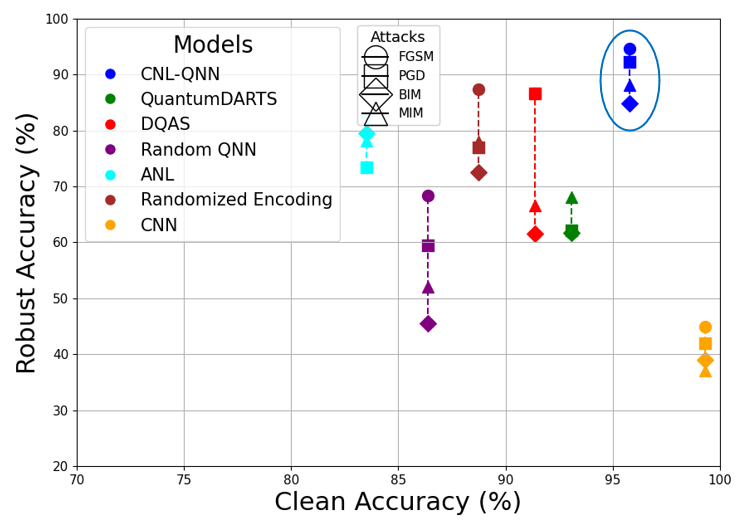} 
\caption{
Comparison of standard and robust accuracy on the MNIST dataset under an $\epsilon=0.3$ attack. The results illustrate the performance of various QNNs and a CNN in both clean conditions and under adversarial attacks (FGSM, PGD, BIM, MIM). The experiments are repeated across five trials with different settings. Our approach achieves a favorable tradeoff between accuracy and robustness for diverse architectures.
}
\label{fig1}
\end{figure}

In this work, we propose a differentiable architecture search framework that incorporates a single \emph{classical noise layer} (CNL) before the quantum circuit. This CNL approach selectively injects mild, trainable perturbations into the input data, preserving high accuracy on clean examples while effectively dampening both adversarial and stochastic fluctuations. Through joint optimization of the QNN architecture and the parameters of the CNL, we achieve robust, high-performance models suitable for challenging vision tasks. Our method is straightforward to implement and avoids the overhead commonly associated with circuit-level quantum defenses. As illustrated in  Figure~\ref{fig1}, experiments on MNIST, Fashion MNIST, and CIFAR clearly demonstrate that our approach consistently improves both robustness and standard accuracy compared to existing baselines.

\section{Related works}
Prior approaches to designing quantum circuits often relied on evolutionary algorithms or reinforcement learning, which can be prohibitively expensive for large-scale applications. To address these challenges, differentiable methods have emerged as a powerful alternative \citepblueyear{zhang2022differentiable, wu2023quantumdarts, chen2024differentiable, sun2023differentiable, martyniuk2024quantum}. By continuously parameterizing gate or connectivity choices, these techniques enable gradient-based optimization over circuit structures and trainable parameters simultaneously, and have proven successful in tasks ranging from image classification to quantum optimization. Although strategies like training-free quantum architecture search exist \citepblueyear{he2024training}, differentiable approaches remain highly appealing for their balance of efficiency and flexibility.

In the classical deep-learning domain, researchers have tackled similar issues by explicitly optimizing architectures for resilience. DSRNA \citepblueyear{hosseini2021dsrna}, for instance, incorporates differentiable robustness metrics like Jacobian norms to guide neural architecture search, and ARNAS \citepblueyear{ou2024towards} employs a multi-objective approach to jointly optimize natural and adversarial losses. Progressive adversarial training methods have also been integrated into architecture search to gradually build robustness during the discovery process. While these strategies achieve notable robustness gains, direct extensions to quantum circuits are hampered by unique noise characteristics and stringent resource constraints.

A crucial concern in quantum neural networks is their vulnerability to adversarial attacks \citepblueyear{maouaki2024robqunns} and noise from near-term hardware \citepblueyear{lu2020quantum}. Several defense mechanisms have been proposed to address these challenges. Quantum noise injection \citepblueyear{huang2023enhancing} introduces controlled noise within quantum circuits to improve stability against perturbations. Randomized encoding techniques \citepblueyear{gong2024enhancing} offer gradient obfuscation to thwart adversarial attacks. Adversarial training methods adapted to quantum settings \citepblueyear{liao2021robust} can enhance robustness by training on perturbed inputs. However, these approaches often introduce significant computational overhead or compromise clean accuracy, creating a challenging trade-off between performance and resilience in practical quantum applications.

While existing DQAS methods focus primarily on optimizing clean accuracy and circuit efficiency, our work extends differentiable architecture search to explicitly address robustness as a core design criterion. Rather than treating robustness as a post-hoc enhancement, CNL-QNN integrates adversarial and noise resilience directly into the architecture search objective. This approach shifts quantum circuit discovery from purely performance-driven optimization to a deployment-oriented framework that accounts for both algorithmic capability and operational stability. By co-optimizing circuit structure and robustness through classical preprocessing, our method discovers architectures that maintain high performance under the practical constraints of near-term quantum hardware, bridging the gap between theoretical circuit design and real-world quantum system deployment.

\section{Problem Statement}
Despite the promise shown by quantum neural networks for computer vision tasks, they remain highly susceptible to both adversarial manipulations and hardware-induced noise. This vulnerability presents a significant barrier to their practical deployment in real-world scenarios. The fundamental issue lies in the existing defense mechanisms: while they improve robustness, they do so at substantial costs:
\begin{itemize}
\item \textbf{Compromised Accuracy:} Methods like quantum noise injection and randomized encoding often lead to degraded performance on clean, unperturbed data.
\item \textbf{Computational Burden:} Adversarial training and complex noise operations within quantum circuits dramatically increase resource requirements, limiting scalability.
\item \textbf{Limited Attack Coverage:} Most prior research addresses only a narrow range of perturbations, leaving models vulnerable to diverse attack vectors.
\end{itemize}

The field currently lacks a solution that can maintain high clean accuracy while providing broad-spectrum robustness without imposing prohibitive computational costs. This gap is particularly problematic when scaling to more complex image classification problems beyond simple datasets, where the computational demands of existing defenses become increasingly impractical. To overcome these limitations, we present a noise-aware architecture search that introduces a single CNL before the quantum circuit. Unlike methods that stack multiple noise gates throughout the circuit, our CNL is both lightweight and trainable, adding minimal overhead while significantly mitigating the model's sensitivity to perturbations. We further integrate this layer into a differentiable quantum architecture search framework \citepblueyear{zhang2022differentiable, wu2023quantumdarts}, jointly optimizing circuit structure and noise parameters to maximize both accuracy and robustness.

Our main contribution is a concise yet effective solution that leverages a single CNL in conjunction with differentiable circuit search to achieve robust quantum models for image classification.

Specifically, our framework:
\begin{itemize}
    \item Maintains high clean accuracy by inserting mild perturbations \emph{before} the quantum circuit, avoiding complex noise operations inside the circuit.
    \item Eliminates the need for resource-intensive adversarial training or multiple quantum noise layers, substantially lowering computational overhead.
    \item Delivers consistent gains on MNIST, Fashion MNIST, and CIFAR, with additional validation on real quantum hardware, demonstrating its clear viability for practical quantum vision applications.
\end{itemize}

\begin{figure*}[t]
\centering
\includegraphics[width=\textwidth]{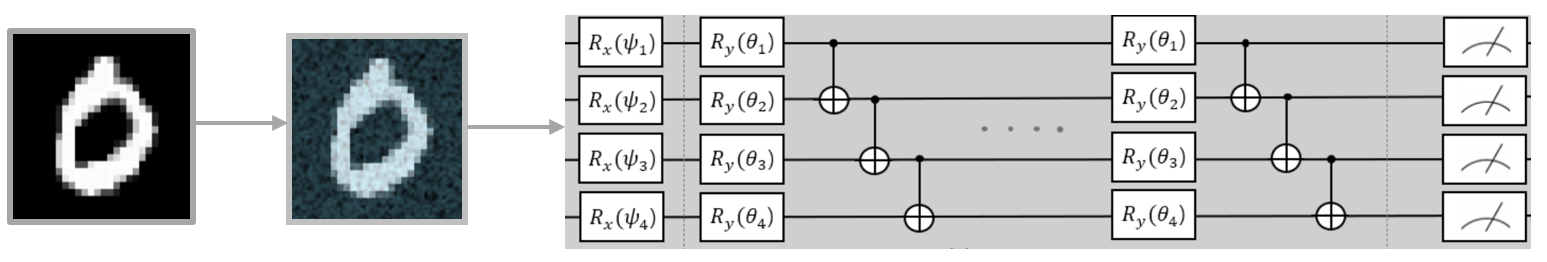}
\caption{Overview of the CNL-QNN framework. A lightweight Classical Noise Layer (CNL) introduces mild, trainable perturbations to the input data before quantum encoding, enhancing robustness without affecting clean accuracy. The workflow proceeds as follows: (1) input preprocessing, where classical images are resized and normalized to match qubit configuration; (2) stochastic perturbation by the CNL, defined as $x' = x + h \cdot \text{sign}(\xi)$, applied at the classical stage; (3) quantum encoding via angle encoding of normalized pixel values into $R_x$ gates; (4) differentiable architecture search using Gumbel-Softmax sampling; and (5) composite optimization through $L_{total} = L_{mse} + \gamma \cdot L_{robust}$. This figure provides a concise high-level view of how classical perturbations and quantum architecture search are jointly optimized to improve robustness without additional quantum resource costs.}
\label{fig2}
\end{figure*}

\section{Quantum Background}
\label{sec:quantum_background}

\subsection{Qubits}
In quantum computing, the fundamental unit of information is the \emph{qubit}, which differs from the classical bit by allowing a superposition of two basis states, typically labeled \(\ket{0}\) and \(\ket{1}\). A general single-qubit state can be written as
\begin{equation}
\ket{\psi} = \alpha \ket{0} + \beta \ket{1},
\end{equation}
where \(\alpha\) and \(\beta\) are complex amplitudes satisfying \(\lvert \alpha \rvert^2 + \lvert \beta \rvert^2 = 1\). This superposition enables the qubit to occupy a continuous range of states between \(\ket{0}\) and \(\ket{1}\), introducing richer possibilities for data representation than classical binary encoding. A convenient way to denote quantum states uses the Dirac or \emph{bra--ket} notation, with \(\langle \cdot \mid \cdot \rangle\) indicating inner products. In vector form, the computational basis states can be expressed as
\[
\ket{0} = 
\begin{pmatrix}
1\\[4pt]
0
\end{pmatrix},
\quad
\ket{1} =
\begin{pmatrix}
0\\[4pt]
1
\end{pmatrix},
\quad
\ket{\psi} =
\begin{pmatrix}
\alpha \\[2pt]
\beta
\end{pmatrix}.
\]

\subsection{Multiple Qubits and Entanglement}
A system of \(n\) qubits resides in a space formed by the tensor product of individual qubit spaces, yielding a \(2^n\)-dimensional state space. For instance, two single-qubit states,
\(\ket{\psi_1} = \alpha\ket{0} + \beta\ket{1}\)
and
\(\ket{\psi_2} = \gamma\ket{0} + \omega\ket{1},\)
combine to form the joint state
\begin{equation}
\ket{\psi_1} \otimes \ket{\psi_2} 
= \alpha \gamma \,\ket{00} 
+ \alpha \omega \,\ket{01}
+ \beta \gamma \,\ket{10}
+ \beta \omega \,\ket{11}.
\end{equation}
Not all multi-qubit states, however, factorize in this way. An \emph{entangled} state cannot be decomposed into separate qubits; a well-known example is the Bell state,
\begin{equation}
\ket{\Phi^{+}} = \frac{1}{\sqrt{2}}\bigl(\ket{00} + \ket{11}\bigr),
\end{equation}
where each qubit’s outcome is correlated with the other upon measurement. Entanglement is crucial in quantum algorithms, as it can enable correlations and transformations unattainable by purely classical means.

\subsection{Measurement and Computation}
To extract classical information from a quantum system, one performs a measurement, typically in the computational basis (or another chosen basis). Measuring the single-qubit state \(\ket{\psi} = \alpha \ket{0} + \beta \ket{1}\) yields the result \(\ket{0}\) with probability \(\lvert \alpha \rvert^2\) and \(\ket{1}\) with probability \(\lvert \beta \rvert^2\). After measurement, the qubit collapses to the observed basis state. Quantum neural networks exploit this framework by preparing a parameterized state, processing it through entangling gates, and then measuring an observable to infer labels or regression targets. These measurements guide updates to the circuit parameters via gradient-based or other learning algorithms.

\section{Methodology}
\label{sec:methodology}

We now detail our proposed \textbf{CNL-QNN} framework, which integrates a single classical noise layer with differentiable architecture search. Figure~\ref{fig2} provides a schematic overview, and Algorithm~\ref{alg:cnlqnn} outlines the training procedure.

\subsection{Search Space and Classical Noise Layer}

Our search space includes both single-qubit rotations (e.g., $R_x$, $R_z$) and common two-qubit gates (CNOT, CRZ, XX, YY, ZZ, CZ, ISWAP), parameterized by weights $\Theta$ and discrete architecture choices $\alpha$. 
Circuit architectures are optimized using Gumbel-Softmax sampling \citepblueyear{jang2016categorical, maddison2016concrete} with temperature annealing, which enables smooth exploration of discrete gate combinations during training. The theoretical foundation for classical noise injection as a robustness mechanism operates through integrated training with perturbed inputs. During the training process, we incorporate a robustness term that evaluates model performance on stochastically perturbed versions of the input data. Specifically, for each training sample $x$, we generate a perturbed version:
\begin{equation}
x_{adv} = x + h \cdot \text{sign}(\xi),
\end{equation}
where $h$ is a small perturbation magnitude (typically $h = 0.02$), $\xi \sim \mathcal{N}(0, I)$ represents Gaussian random noise, and $\text{sign}(\cdot)$ extracts the sign of each component. The total training loss combines the standard mean squared error with a robustness term:
\begin{equation}
L_{total} = L_{mse}(y, \hat{y}) + \gamma \cdot L_{robust}(y, \hat{y}_{adv}),
\end{equation}
where $\gamma$ (typically set to 1) controls the balance between clean accuracy and robustness, $\hat{y}$ is the prediction on clean input, and $\hat{y}_{adv}$ is the prediction on the perturbed input.

This mechanism enhances robustness through two complementary effects. First, by explicitly training on randomly perturbed inputs, the quantum circuit learns feature representations that are inherently less sensitive to small input variations. The model must maintain correct classifications despite these perturbations, effectively smoothing the decision boundary in regions where adversarial attacks might otherwise succeed. Second, the stochastic perturbations act as implicit data augmentation within the quantum feature space. Each training sample is effectively presented to the quantum circuit in multiple slightly varied forms throughout training, increasing the diversity of quantum states encountered without requiring additional quantum measurements. Unlike explicit data augmentation that would necessitate multiple quantum circuit executions per sample, our approach achieves similar regularization benefits at the classical preprocessing stage.

The computational efficiency of this approach stems from its classical nature. The perturbation operation $x_{adv} = x + h \cdot \text{sign}(\xi)$ requires only $O(d)$ operations where $d$ is the input dimension, which is negligible compared to the $O(2^n \cdot L)$ complexity of quantum circuit execution with $n$ qubits and depth $L$. Empirically, incorporating the robustness term adds less than 2\% to total training time per epoch. Critically, unlike quantum noise injection or randomized encoding methods that require additional quantum gates and measurements within the quantum circuit, our classical preprocessing consumes zero quantum resources. This makes the approach particularly suitable for near-term quantum devices where quantum operations are expensive and quantum coherence time is limited. The robustness benefits are substantial, as demonstrated in our experiments where removing this mechanism results in 12-14\% accuracy degradation under adversarial attacks while maintaining it preserves both clean and robust accuracy.

\subsection{Differentiable Sampling}
To handle the discrete nature of gate selection, we adopt the Gumbel-Softmax technique, which approximates a categorical choice by a continuous, differentiable sample. Concretely, for a search space of \(k\) gate types, we maintain a probability vector \(\pi \in \mathbb{R}^k\). Sampling a gate \(i\) from \(\pi\) is relaxed into
\[
y_i = \frac{\exp\bigl((\log(\pi_i) + g_i) / \tau\bigr)}
  {\sum_{j=1}^k \exp\bigl((\log(\pi_j) + g_j) / \tau\bigr)},
\]
where \(g_i\) are Gumbel(0,1) samples and \(\tau\) is a temperature hyperparameter. 
\begin{algorithm}[tb]
\caption{CNL-QNN Architecture Search}
\label{alg:cnlqnn}
\textbf{Input}: Dataset $(x, y)$, number of layers $p$, operation pool $\mathcal{G}$, number of sampled architectures per epoch $n_{\text{arch}}$, number of iterations per epoch $n_{\text{iter}}$, batch size $K$, total epochs $T$\\
\textbf{Parameters}: Architecture logits $\alpha \in \mathbb{R}^{p \times |\mathcal{G}|}$, circuit parameters $\omega$, learning rates $\mathrm{lr}_{\omega}, \mathrm{lr}_{\alpha}$, temperature $\tau$\\
\textbf{Output}: Optimal architecture $k^*$ and final parameters $\omega^*$

\begin{algorithmic}
\State Initialize $\alpha$ and $\omega$
\For{epoch $= 1$ to $T$}
    \For{architecture index $= 1$ to $n_{\text{arch}}$}
        \State Sample a circuit architecture $k$ using Gumbel-Softmax with logits $\alpha$ and temperature $\tau$
        \For{iteration $= 1$ to $n_{\text{iter}}$}
            \State Sample a mini-batch $(x_j, y_j)$ of size $K$
            \State Apply the classical noise layer to $x_j$
            \State Compute model outputs $f_{k}(x_j; \omega)$
            \State Compute training loss $\mathcal{L}_{\text{train}}(f_{k}(x_j; \omega), y_j)$ 
            \State (Optionally) incorporate robustness objectives into a total loss $\mathcal{L}_{\text{total}}$
            \State Compute $\nabla_\omega \mathcal{L}_{\text{total}}$ and $\nabla_\alpha \mathcal{L}_{\text{total}}$
            \State Update circuit parameters: $\omega \leftarrow \omega - \mathrm{lr}_{\omega} \,\nabla_\omega \mathcal{L}_{\text{total}}$
            \State Accumulate gradients for architecture: $\nabla_\alpha \mathcal{L}_{\text{total}}$
        \EndFor
    \EndFor
    \State Update $\alpha \leftarrow \alpha - \mathrm{lr}_{\alpha} \,\nabla_\alpha \mathcal{L}_{\text{total}}$ 
\EndFor
\State Select best architecture $k^*$ using validation accuracy
\State \textbf{return} $k^*$ and $\omega^*$
\end{algorithmic}
\end{algorithm}

\subsection{Objective Function}
We aim to jointly optimize \(\Theta\) (circuit parameters and noise-layer weights) and \(\alpha\) (architecture decisions). Let \(\mathcal{L}_{\mathrm{train}}(\Theta,\alpha)\) be the standard loss on clean data, and let \(\mathcal{L}_{\mathrm{robust}}(\Theta,\alpha)\) account for performance under adversarial or hardware noise scenarios. Our total objective is
\[
\mathcal{L}_{\mathrm{total}}(\Theta,\alpha) 
= \mathcal{L}_{\mathrm{train}}(\Theta,\alpha) 
+ \lambda \,\mathcal{L}_{\mathrm{robust}}(\Theta,\alpha),
\]
where \(\lambda\) controls the trade-off between standard accuracy and robustness. In practice, \(\mathcal{L}_{\mathrm{robust}}\) can incorporate adversarial samples generated via standard attack algorithms (e.g.\ FGSM, PGD) or realistic quantum noise models.

\noindent\textbf{Training Procedure.}
Algorithm~\ref{alg:cnlqnn} outlines the overall workflow of our CNL-QNN approach. We maintain two sets of parameters: (1) \(\alpha\), which governs the discrete gate choices in the circuit architecture, and (2) \(\omega\), which includes both the continuous gate parameters and any adjustable terms for the classical noise layer. At each epoch, we sample multiple candidate architectures using the Gumbel-Softmax relaxation with \(\alpha\). For each candidate, we iterate several parameter-update steps: drawing a mini-batch of data, applying the classical noise layer to inputs, and computing the forward pass through the chosen circuit. We combine the standard training loss with any optional robustness objectives (e.g.\ adversarially perturbed samples, noise models) to obtain the total loss. The circuit parameters \(\omega\) are updated after each mini-batch, while gradients with respect to \(\alpha\) are accumulated and used to update \(\alpha\) once per epoch. After training, we select the best-performing architecture based on validation accuracy and finalize its parameters.

\section{Experiments}

\subsection{Quantum Model Selection Criteria}
In our study, we established rigorous criteria for model selection to ensure a comprehensive and fair evaluation of quantum architectures. We focused on differentiable search methods due to their significant efficiency advantages over traditional approaches like reinforcement learning and evolutionary algorithms \citepblueyear{zhang2022differentiable}. The high computational demands of evolutionary algorithms and reinforcement learning often render them impractical for extensive architecture exploration, particularly in quantum models where computational resources are inherently limited. In contrast, DQAS approaches offer an effective balance between thorough search space exploration and computational efficiency, making them our preferred methodology for this investigation.

\begin{figure}[t]
    \centering
    \includegraphics[width=1\columnwidth]{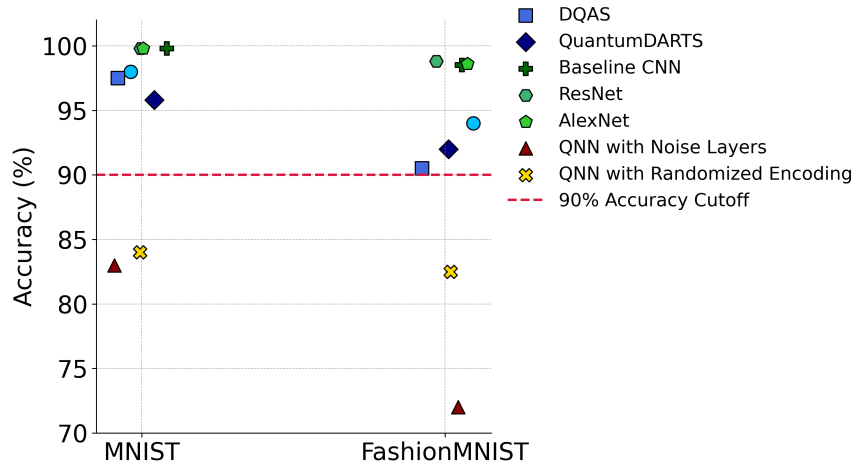}
    \caption{Accuracy comparison on a 9-qubit configuration, tested for MNIST and FashionMNIST. Models such as a QNN with an internal quantum noise layer and one using randomized encoding fell below the 90\% accuracy cutoff, indicating limited clean-performance on these datasets. For CIFAR, the performance across quantum models was too close to be a useful discriminator.}
    \label{fig:model_accuracy_comparison}
\end{figure}

Our selection process involved systematic preliminary evaluations on multiple quantum model variants to determine which architectures warranted further investigation. Figure \ref{fig:model_accuracy_comparison} illustrates the performance comparison across different quantum models on a standardized 9-qubit configuration, tested on both MNIST and FashionMNIST datasets. We established a minimum performance threshold of 90\% accuracy on clean data as an inclusion criterion for subsequent robustness testing. This threshold represents a reasonable balance between model capability and practical utility for real-world applications.

Several robustness-enhancing approaches were initially considered but ultimately excluded based on their failure to meet our performance criteria. Specifically, QNNs incorporating internal quantum noise layers and those utilizing randomized encoding techniques consistently fell below our accuracy threshold. This observation aligns with prior research demonstrating that while these approaches can enhance robustness, they often do so at the expense of clean data performance \citepblueyear{huang2023enhancing, gong2024enhancing}. The significant accuracy degradation observed in these models (often 5-8\% below our baseline QNN) made them unsuitable candidates for applications requiring both robustness and high accuracy.

To establish meaningful baselines for comparison, we included a Random QNN with arbitrarily selected gate configurations. This baseline serves as an important control to quantify the effectiveness of structured search methods against randomly assembled quantum circuits, providing insight into whether performance gains result from sophisticated search strategies or simply from the quantum computing paradigm itself. The substantial performance gap observed between random architectures and those discovered through differentiable search (typically 7-12\% on MNIST) validates the importance of principled architecture optimization in the quantum domain.

For classical comparisons, we evaluated several established neural network architectures, including ResNet and AlexNet variants, alongside a standard CNN baseline. Interestingly, these advanced classical architectures showed negligible performance improvements over the baseline CNN on our target datasets, likely due to the relative simplicity of these datasets, which do not fully leverage the increased capacity of complex classical models. Consequently, we retained the baseline CNN as our representative classical model.

\subsection{Experimental Setup}
To evaluate the effectiveness of the CNL-QNN framework, we conducted experiments with binary classification on benchmark datasets, including MNIST, Fashion MNIST, and CIFAR, the last two are rarely explored in quantum models due to their complexity. Current QNNs are not yet capable of processing large-scale datasets like ImageNet or other benchmarks commonly used in classical computer vision, which limits evaluations to smaller vision tasks. Input images are resized to match the available qubit configuration: $2 \times 2$ grids for 4-qubit systems, $3 \times 3$ grids for 9-qubit systems, and $4 \times 4$ grids for 16-qubit systems. Pixel values are normalized from the range [0, 255] to $[0, 2\pi]$ using the transformation $x_{normalized} = x_{raw} \times 2\pi / 255.0$. These normalized values serve as rotation angles for angle encoding, where each pixel value directly parameterizes an $R_x$ gate applied to its corresponding qubit. This encoding scheme preserves feature variance while enabling fair comparison across different qubit counts. These tests aimed to assess the performance and robustness of the discovered QNN architectures under various conditions, including adversarial attacks and quantum noise generated by 3 different models. The experiments, performed using TensorFlow Quantum on an NVIDIA A100 GPU, utilized quantum circuits with XX, YY, ZZ, CZ, CNOT, and ISWAP gates. Stochastic Gradient Descent (SGD) optimized model weights with learning rate 0.01, while the Adam optimizer tuned architecture parameters $\alpha$ with learning rate 0.01. Training used batch size 32, with 3 sampled architectures per epoch during the search phase. Gumbel-Softmax temperature was initialized at $\tau_0 = 5.0$ and decayed by a factor of 0.95 per epoch. Architecture search ran for 5-10 epochs with early stopping based on validation performance (patience 5), followed by 5-10 epochs of final training on the discovered architecture.

The Classical Noise Layer operates by applying stochastic perturbations to input data before quantum encoding. For each training sample $x$, we generate perturbed versions using $x_{adv} = x + h \cdot \text{sign}(\xi)$, where $h = 0.02$ is the perturbation magnitude and $\xi \sim \mathcal{N}(0, I)$ represents Gaussian random noise sampled independently at each training step. The parameter $h$ is a single scalar hyperparameter determined through initial validation and held fixed throughout training, adding $O(1)$ parameters to the overall framework. The robustness weight $\gamma = 1$ balances the standard classification loss with the robustness term in $L_{total} = L_{mse}(y, \hat{y}) + \gamma \cdot L_{robust}(y, \hat{y}_{adv})$. Gradients for quantum circuit parameters are computed using TensorFlow Quantum's parameter-shift rule, which evaluates circuit outputs at shifted parameter values to estimate derivatives. For the classical preprocessing parameters (when trainable), gradients are obtained through standard automatic differentiation. This design requires no additional quantum operations, as all perturbations occur at the classical preprocessing stage before quantum encoding. The implementation uses standard TensorFlow operations without relying on external noise injection APIs or specialized quantum noise libraries.

We compared CNL-QNN against QuantumDARTS \citepblueyear{wu2023quantumdarts}, DQAS \citepblueyear{zhang2022differentiable}, a Random QNN with randomly selected gates, and a classical CNN to understand their relative vulnerabilities to adversarial attacks. This comparative analysis is crucial for evaluating how quantum models stack up against classical ones in terms of robustness.

\subsection{Performance Against Adversarial Attacks}
Robustness against adversarial attacks remains a critical concern for QNNs in image classification, where even small, carefully crafted perturbations can trigger misclassifications. To assess this vulnerability, standard attack methods like Fast Gradient Sign Method (FGSM), Projected Gradient Descent (PGD), Basic Iterative Method (BIM), and Momentum Iterative Method (MIM) are applied, each generating deceptive inputs by exploiting the model’s gradients. In QNNs, however, collecting gradient information is often more complex due to parameter-shift rules, measurement overhead, and the non-linear effects of entanglement. Consequently, adversarial attacks tend to be more computationally expensive and less straightforward to implement, yet they can still undermine QNN performance if defenses are inadequate. As a result, designing architectures that inherently reduce sensitivity to such perturbations is paramount for secure and reliable quantum computing in real-world settings.

\begin{algorithm}[!t]
\caption{Quantum PGD Attack}
\label{alg:pgd}
\begin{algorithmic}[1] 
\State Initialize adversarial example $x_{\text{adv}} = x$ and input qubits
\For{each sample $(x_i, y_i)$}
    \State Set $x_{\text{adv}_i} = x_i$
    \For{step $s = 1$ to $n$ iterations}
        \State Flatten and concatenate $x_{\text{adv}_i}$ with model weights
        \State Compute model output $y_{\text{pred}}$ using $x_{\text{adv}_i}$ and quantum circuit
        \State Calculate loss $\mathcal{L}$ between $y_{\text{pred}}$ and $y_i$
        \State Compute gradients $\nabla_{x_{\text{adv}_i}} \mathcal{L}$
        \State Update $x_{\text{adv}_i} \gets x_{\text{adv}_i} + \epsilon \cdot \text{sign}(\nabla_{x_{\text{adv}_i}} \mathcal{L})$
        \State Project $x_{\text{adv}_i}$ onto the $\ell_\infty$ ball around $x_i$
    \EndFor
    \State Set $x_{\text{adv}}[i] = x_{\text{adv}_i}$
\EndFor
\State \textbf{return} $x_{\text{adv}}$
\end{algorithmic}
\end{algorithm}

\begin{table*}[t]
\centering
\caption{Performance of models on MNIST, Fashion MNIST, and CIFAR Datasets under various attacks.}
\label{table:combined-performance}
\begin{tabular}{l|c|c|c|c|c}
    Model & Clean Accuracy & FGSM (0.3) & PGD (0.3) & BIM (0.3) & MIM (0.3) \\
    \hline
    \multicolumn{6}{c}{MNIST} \\
    \hline
    CNL-QNN (Ours) & 97.97$\pm$0.4 & \textbf{91.84$\pm$2.7} & \textbf{88.72$\pm$3.1} &\textbf{84.82$\pm$2.0} & \textbf{88.14$\pm$0.6} \\
    QuantumDARTS & 95.66$\pm$3.3 & 63.44$\pm$8.9 & 84.58$\pm$0.8 & 63.71$\pm$6.1 & 62.15$\pm$2.5 \\
    DQAS & 96.28$\pm$3.9 & 85.50$\pm$6.6 & 84.76$\pm$8.1 & 61.64$\pm$12.46 & 58.65$\pm$12.5 \\
    Random QNN & 63.12$\pm$21.2 & 56.23$\pm$33.2 & 61.42$\pm$23.5 & 65.52$\pm$14.1 & 55.12$\pm$13.8 \\
    CNN & \textbf{99.32$\pm$0.04} & 61.83$\pm$0.04 & 14.32$\pm$1.3 & 38.14$\pm$0.5 & 31.00$\pm$0.2 \\
    \hline
    \multicolumn{6}{c}{Fashion MNIST} \\
    \hline
    CNL-QNN (Ours) & 93.86$\pm$1.7 & \textbf{83.2$\pm$2.9} & \textbf{76.5$\pm$5.6} & \textbf{78.2$\pm$5.3} & \textbf{79.5$\pm$3.3} \\
    QuantumDARTS & 92.2$\pm$0.06 & 73.0$\pm$1.1 & 68.5$\pm$5.6 & 69.6$\pm$1.7 & 59.2$\pm$8.4 \\
    DQAS & 93.40$\pm$0.09 & 75.4$\pm$3.4 & 68.1$\pm$6.8 & 74.6$\pm$3.6 & 74.0$\pm$1.1 \\
    Random QNN & 77.14$\pm$10.1 & 478.0$\pm$6.4 & 63.5$\pm$11.6 & 35.2$\pm$23.9 & 36.0$\pm$16.9 \\
    CNN & \textbf{99.03$\pm$0.04} & 13.60$\pm$0.8 & 6.12$\pm$2.0 & 47.4$\pm$0.4 & 41.5$\pm$1.3 \\
    \hline
    \multicolumn{6}{c}{CIFAR} \\
    \hline
    CNL-QNN (Ours) & 71.82$\pm$1.6 & \textbf{66.4$\pm$4.8} & \textbf{64.8$\pm$4.0} & \textbf{61.2$\pm$7.8} & 59.8$\pm$2.6 \\
    QuantumDARTS & 70.14$\pm$1.8 & 61.6$\pm$1.6 & 64.6$\pm$0.8 & 60.8$\pm$5.0 & \textbf{62.20$\pm$8.2} \\
    DQAS & 71.12$\pm$5.7 & 63.66$\pm$1.0 & 64.5$\pm$0.4 & 61.2$\pm$2.8 & 61.4$\pm$2.6 \\
    Random QNN & 66.20$\pm$9.4 & 54.40$\pm$3.0 & 33.4$\pm$10.4 & 48.2$\pm$12.2 & 42.8$\pm$14.4 \\
    CNN & \textbf{95.40$\pm$0.03} & 29.4$\pm$2.0 & 18.4$\pm$2.3 & 31.8$\pm$0.8 & 21.0$\pm$1.1 \\
\end{tabular}
\end{table*}

FGSM is a single-step attack that perturbs input data in the direction of the gradient of the loss function \citepblueyear{huang2017adversarial, ozdag2018adversarial}. While straightforward in classical networks, applying FGSM in QNNs involves more complexity due to the quantum nature of the model parameters. The computation of gradients in quantum circuits requires careful consideration of quantum gates and measurements, making the attack generation process more intricate.
PGD, an iterative extension of FGSM, refines this approach by repeatedly applying perturbations and projecting the perturbed inputs back onto a constrained space \citepblueyear{villegas2024evaluating,majumder2021hybrid}. as outlined in Algorithm \ref{alg:pgd}. This method allows for more fine-tuned adversarial examples but also introduces additional complexity in quantum models. 

The performance of models on MNIST, Fashion MNIST, and CIFAR datasets under various attacks is summarized in Table \ref{table:combined-performance}, where CNL-QNN consistently demonstrates strong performance even under high perturbation levels.

\begin{figure}[t]
\centering
 \includegraphics[width=1\columnwidth]{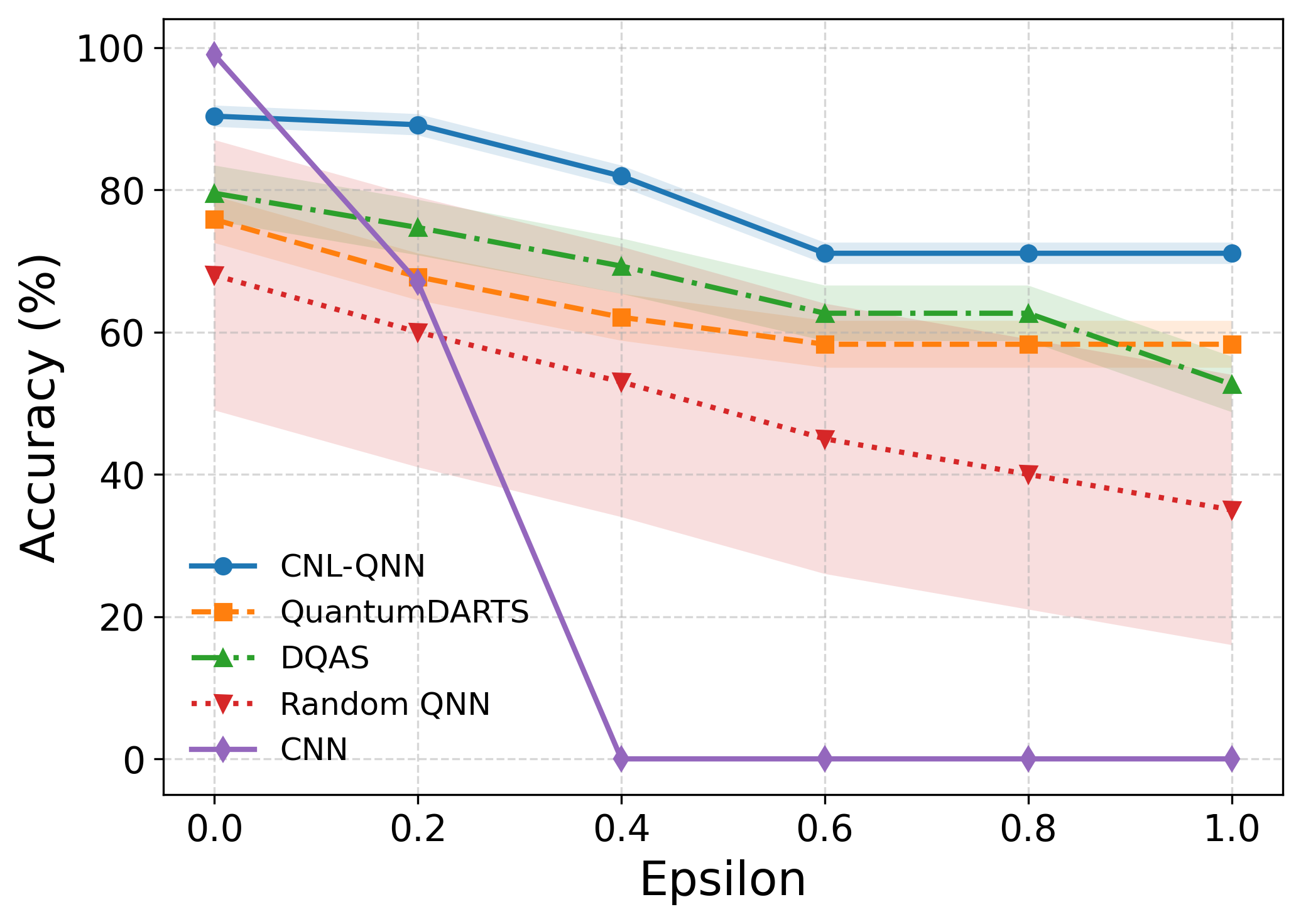} 
\caption{Performance of different quantum models under PGD attack with 4 qubits. The CNL-QNN model exhibits a slight degradation in performance as epsilon increases but maintains higher robustness compared to other QNNs. The Random QNN and CNN models show a significant drop in accuracy under higher epsilon values, indicating their susceptibility to adversarial attacks. The shading in the figure represents the standard deviation across the 5 experiments.}
\label{fig4}
\end{figure}

\begin{figure}[t]
\centering
\includegraphics[width=1\columnwidth]{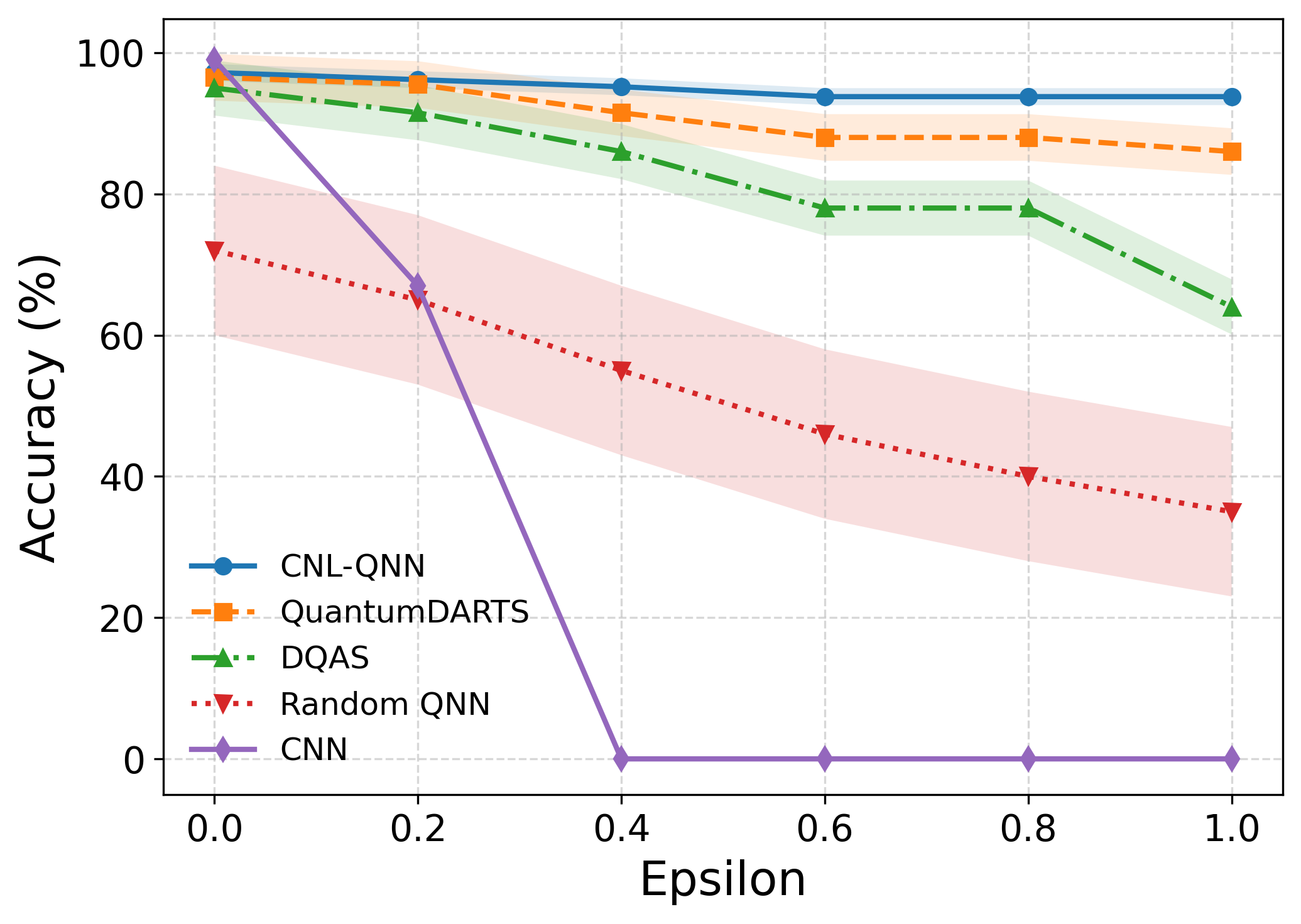} 
\caption{Performance of different quantum models under PGD attack with 9 qubits. QNNs with 9 qubits generally demonstrate increased robustness compared to models with fewer qubits. The CNL-QNN model, along with other QNNs, maintains higher accuracy across varying epsilon values, highlighting the advantage of higher qubit counts in improving resistance to adversarial attacks.}
\label{fig5}
\end{figure}

Model complexity played a crucial role in robustness, with nine-qubit models showing significantly better resilience to attacks than four-qubit models. The increased qubit count provided greater capacity, enhancing the model's ability to withstand adversarial perturbations. Figures \ref{fig4} and \ref{fig5} compare the performance of different models under PGD attacks using 4 qubits and 9 qubits, respectively. As the number of qubits increases, most QNNs generally become more robust and exhibit improved performance. In contrast, while CNNs outperform QNNs in clean accuracy, they are significantly more sensitive to adversarial attacks.

\subsection{Performance Under Circuit Noises}
In our experiments, quantum noise was integrated within the circuits, reflecting the complexities encountered in real quantum systems. For depolarizing noise, we introduced randomness during the application of quantum gates. This process involved randomly applying Pauli gates (X, Y, or Z) to the qubits during the execution of key operations in our search space. This noise affected both single-qubit gates and two-qubit operations like CNOT and ISWAP. In the case of these two-qubit gates, the depolarizing noise not only impacted individual qubits but also the entangled states created by these gates, simulating the real-world scenario where entanglement can be disturbed, leading to errors in the overall quantum state \citepblueyear{patterson2021quantum, dur2005standard}

For phase flip noise, random phase alterations were introduced during or after applying gates like XX, ZZ, and CZ, simulating disruptions in phase coherence crucial for these operations \citepblueyear{schindler2011experimental}. Bit flip noise was integrated by introducing random flips during circuit execution, affecting both individual qubits and the fidelity of operations like CNOT and ISWAP \citepblueyear{xue2021effects}. This approach allowed us to observe how these types of noise impacted the reliability of quantum operations under realistic conditions.

\begin{figure}[t]
\centering
\includegraphics[width=1\columnwidth]{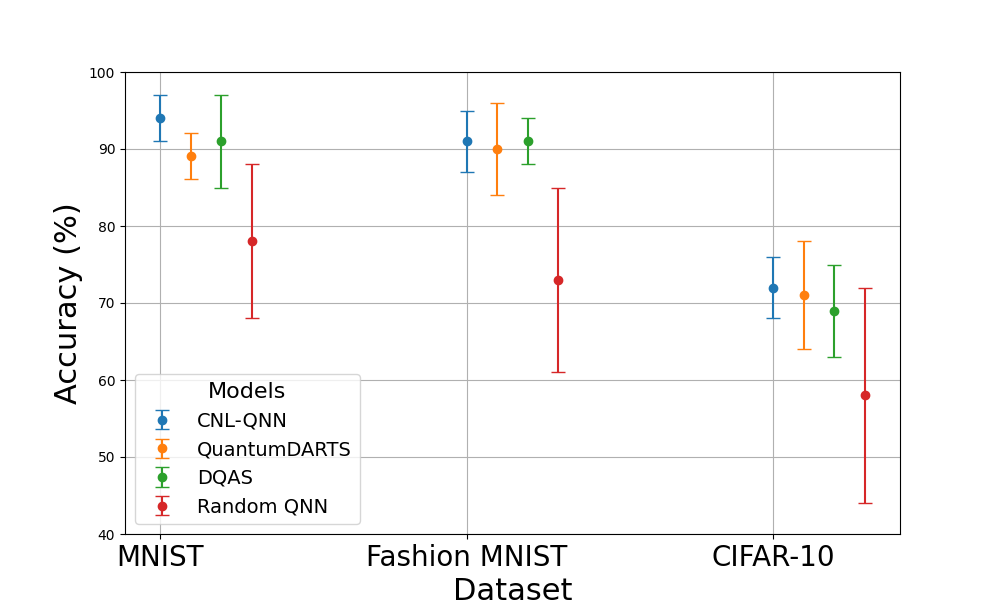} 
\caption{Model performance across the 3 datasets, evaluated under three noise models (phase flip, bit flip, and depolarizing noise) with varying noise levels (5\%, 8\%, and 10\%). The accuracy is represented by the points, and the error bars indicate the variance in model performance across the different noise levels. CNL-QNN generally demonstrates higher accuracy and lower variance compared to other QNNs.}
\label{fig6}
\end{figure}

Figure \ref{fig6} illustrates the model performance across the 3 datasets, evaluated under three noise models, with varying noise levels (5\%, 8\%, and 10\%). The accuracy is represented by the points, and the error bars indicate the variance in model performance across the different noise levels. CNL-QNN generally demonstrates higher accuracy and lower variance compared to QuantumDARTS, DQAS, and Random QNN. Notably, the four QNN models maintained their relative order, with the average accuracy decreasing by 3.8\% compared to the noiseless models with the same parameters.

\subsection{Performance Under Black-Box Attacks}
Black-box attacks refer to scenarios where the adversary does not have access to the model's parameters or architecture but can query the model to generate adversarial examples \citepblueyear{li2022transferability, gong2022universal}. We simulated black-box attacks using both FGSM and PGD with an epsilon value of 0.3. An external CNN was employed to generate the perturbed images. The results, as shown in Table \ref{table:black-box-performance}, demonstrate that the QNNs exhibit increased robustness against black-box attacks compared to their white-box counterparts.

\begin{table}[h]
\centering
\caption{Performance of models under black-box attacks (FGSM and PGD, $\epsilon=0.3$).}
\label{table:black-box-performance}
\begin{tabular}{l|c|c}
    Model & FGSM (0.3) & PGD (0.3) \\
    \hline
    CNL-QNN (Ours) & \textbf{95.62$\pm$2.3} & \textbf{90.52$\pm$6.6} \\
    QuantumDARTS & 84.64$\pm$0.7 & 81.50$\pm$1.2 \\
    DQAS & 86.42$\pm$4.6 & 76.24$\pm$8.2 \\
    Random QNN & 61.40$\pm$12.8 & 28.66$\pm$3.3 \\
    CNN & 28.12$\pm$0.4 & 4.60$\pm$3.2 \\
\end{tabular}
\end{table}

\subsection{Implementation on Real Quantum Hardware}

We extended our evaluation to real quantum hardware using the IBM Sherbrooke backend via Qiskit’s Runtime Service. Given the constraints of transferring our TensorFlow Quantum-based framework directly to Qiskit, we focused on testing the robustness of three discovered quantum architectures: CNL-QNN, QuantumDARTS, and DQAS, each configured with the COBYLA optimizer for 20 iterations. Throughout these experiments on the MNIST dataset, we applied a learning rate of 0.01 and a 20\% validation ratio, and we employed Qiskit’s neural network classifier to ensure compatibility with the EstimatorQNN framework. An optimization level of 3 and a resilience level of 2 were maintained for error mitigation. As shown in Table \ref{table:real-quantum-performance}, the CNL-QNN model achieved competitive accuracy, while QuantumDARTS and DQAS followed with slightly lower average accuracies. These reduced values compared to noiseless simulations likely stem from the increased noise and error rates inherent in real quantum hardware.

\begin{table}[h]
\centering
\caption{Performance of quantum models on a real quantum computer using IBM's Sherbrooke backend, across different qubit counts.}
\label{table:real-quantum-performance}
\begin{tabular}{l|c|c|c|c}
    Qubit Count & CNL (Ours) & QDARTS & DQAS & Random \\
    \hline
    4 qubits & \textbf{65.34 }& 63.72 & 61.98 & 64.11 \\
    \hline
    9 qubits &\textbf{ 88.12} & 87.25 & 76.48 & 78.65 \\
    \hline
    16 qubits & \textbf{95.45} & 93.68 & 90.31 & 88.12 \\
\end{tabular}
\end{table}

\subsection{Ablation Studies}
\begingroup\color{black}
We evaluated the contribution of the CNL to robustness by systematically removing it from the framework, effectively reducing CNL-QNN to a DQAS-based architecture search model. This ablation yields performance characteristics comparable to the standard DQAS implementation, allowing for direct assessment of the CNL's impact on robustness while maintaining equivalent circuit discovery capabilities. To select an evaluation strength, we tested multiple $\epsilon$ values and found that $\epsilon<0.2$ produced attacks that were too weak for meaningful assessment, while $\epsilon>0.5$ yielded visually detectable perturbations outside realistic threat models. We therefore used $\epsilon=0.3$ for detailed analysis.Under FGSM at $\epsilon=0.3$ on MNIST and FashionMNIST, removing the CNL reduced robust accuracy by 14.2\% for 4 qubit configurations and by 7.4\% for 9 qubit configurations, averaged across five trials. Paired t tests confirmed significance ($p<0.01$ for 4 qubits, $p<0.05$ for 9 qubits).

We also measured computational overhead using five independent runs under consistent data settings (2000 training and 500 test samples). Table~\ref{tab:training_time} reports average training times for quantum models across 4, 9, and 16 qubits. These measurements show that quantum circuit evaluation and search dominate runtime and that the CNL contributes only a small preprocessing cost while delivering consistent robustness improvements.

\begin{table}[h]
\centering
\caption{Average training time in minutes for quantum models across qubit settings. A classical CNN trained on the same data averaged 6.4 minutes.}
\label{tab:training_time}
\begin{tabular}{lccc}
\hline
\textbf{Model} & \textbf{4 Qubits} & \textbf{9 Qubits} & \textbf{16 Qubits} \\
\hline
CNL QNN (Ours) & 4.2 & 15.6 & 22.4 \\
CNL QNN (w/o CNL) & 3.4 & 14.2 & 21.3 \\
Random QNN & 2.3 & 8.3 & 18.0 \\
DQAS & 3.8 & 16.3 & 23.9 \\
QuantumDARTS & 3.6 & 10.7 & 19.6 \\
\hline
\end{tabular}
\end{table}

As summarized in Table~\ref{tab:training_time}, training time increases with qubit count and search complexity. The CNL adds a small, stable cost relative to quantum evaluation while improving robustness, which supports practicality on near term systems.
\endgroup

\section{Discussion and Limitations}
The results from our experiments demonstrate that CNL-QNN effectively enhances the robustness and performance of QNNs. By integrating CNL with QNNs consistently achieves superior clean accuracy and robustness against adversarial attacks, including FGSM, PGD, BIM, and MIM, across multiple datasets. Our experiments under noisy conditions, including depolarizing, bit flip, and phase flip noise, further validate the resilience of architectures discovered by CNL-QNN. These results indicate that CNL-QNN is not only capable of finding high-performing QNN architectures but also those inherently more resilient to the challenges posed by real-world quantum hardware, which often suffers from noise and other quantum effects. While CNL-QNN significantly advances the design of robust, high-performing QNNs, the introduction of noise adds a slight computational cost. This additional cost is minimal relative to the enhanced robustness achieved, and it remains far more efficient than resource-intensive methods, thereby making this a highly efficient trade-off for the robustness benefits provided.

Looking forward, the scalability of our approach to larger datasets is primarily limited by available quantum hardware capacity rather than the CNL-QNN framework itself. Current quantum devices restrict the size and complexity of datasets that can be effectively processed, limiting applications to smaller-scale datasets. As quantum hardware capabilities continue to improve with more available qubits and enhanced processing power, our framework can naturally scale to accommodate more complex datasets. Promising future directions include developing hierarchical quantum encoding schemes and exploring hybrid decomposition methods that distribute computational load between classical and quantum processors.

\section{Conclusion}
We introduced CNL-QNN, a framework that pairs a single classical noise layer with differentiable architecture search to bolster the robustness of quantum neural networks. By injecting mild perturbations at the input stage, it effectively preserves clean accuracy while mitigating vulnerabilities to adversarial and hardware-induced fluctuations, without the computational burden of previous approaches.

Our experiments on MNIST, Fashion MNIST, and CIFAR demonstrate that this approach consistently discovers architectures resilient under various attacks and noise conditions. This strong performance arises from the synergy between classical noise and gradient-based circuit exploration, offering a practical means to strengthen near-term quantum computing systems. Future extensions include adapting CNL-QNN to broader quantum learning tasks, refining the noise layer for improved hardware compatibility, and examining its impact under more diverse adversarial scenarios.

\section{Acknowledgment}
This work was supported in part by the National Science Foundation under Grant 2335788 and Grant 2343535.

\bibliographystyle{unsrtnat}
\bibliography{references}   

\end{document}